\documentclass[prc,twocolumn,showpacs,preprintnumbers,amsmath,amssymb,superscriptaddress,floatfix,nofootinbib]{revtex4}
\usepackage{graphicx}
\usepackage{amsmath}
\usepackage{amsfonts}
\usepackage{amssymb}%
\usepackage[dvips]{color}

\newcommand{\Slash}[1]{\ooalign{\hfil/\hfil\crcr$#1$}}

\begin{document}

\title{Role of the possible $\Sigma^*(\frac{1}{2}^-)$ state in the $\Lambda p \to \Lambda p \pi^0$ reaction}
\author{Ju-Jun Xie} \email{xiejujun@impcas.ac.cn}
\affiliation{Institute of Modern Physics, Chinese Academy of
Sciences, Lanzhou 730000, China} \affiliation{Research Center for
Hadron and CSR Physics, Institute of Modern Physics of CAS and
Lanzhou University, Lanzhou 730000, China} \affiliation{State Key
Laboratory of Theoretical Physics, Institute of Theoretical Physics,
Chinese Academy of Sciences, Beijing 100190, China}
\author{Jia-Jun Wu}
\affiliation{Physics Division, Argonne National Laboratory, Argonne,
Illinois 60439, USA}
\author{Bing-Song Zou} \email{zoubs@itp.ac.cn}
\affiliation{State Key Laboratory of Theoretical Physics, Institute
of Theoretical Physics, Chinese Academy of Sciences, Beijing 100190,
China}

\begin{abstract}

The $\Lambda p \to \Lambda p \pi^0$ reaction near threshold is
studied within an effective Lagrangian method. The production
process is described by single-pion and single-kaon exchange. In
addition to the role played by the $\Sigma^*(1385)$ resonance of
spin-parity $J^P = 3/2^+$, the effects of a newly proposed
$\Sigma^*$ ($J^P = 1/2^-$) state with mass and width around $1380$
MeV and $120$ MeV are investigated. We show that our model leads to
a good description of the experimental data on the total cross
section of the $\Lambda p \to \Lambda p \pi^0$ reaction by including
the contributions from the possible $\Sigma^*(\frac{1}{2}^-)$ state.
However, the theoretical calculations by considering only the
$\Sigma^*(1385)$ resonance fail to reproduce the experimental data,
especially for the enhancement close to the reaction threshold. On
the other hand, it is found that the single-pion exchange is
dominant. Furthermore, we also demonstrate that the angular
distributions provide direct information of this reaction, hence
could be useful for the investigation of the existence of the
$\Sigma^*(\frac{1}{2}^-)$ state and may be tested by future
experiments.

\end{abstract}

 \maketitle

\section{Introduction}

Study of the spectrum of the $\Sigma(1193)$ excited states,
$\Sigma^*$, with isospin $I=1$ and strangeness $S=-1$ is one of the
most important issues in hadronic
physics~\cite{Klempt:2009pi,Crede:2013kia}. The $\Sigma^*$
resonances were mostly produced and studied in $K$-induced
reactions. Many $\Sigma^*$ resonances are now cataloged in the
particle data group (PDG)~\cite{pdg2012}. However, our knowledge on
these resonances is still very
poor~\cite{Klempt:2009pi,Crede:2013kia,pdg2012}. In the energy
region below $2$ GeV, only a few of them are well established, such
as the $\Sigma^*(1385)$ of spin-parity $J^P = 3/2^+$,
$\Sigma^*(1670)$ of $J^P = 3/2^-$ and $\Sigma^*(1775)$ of $J^P
=5/2^-$. The others are not well established with some even of large
uncertainties on their existence. Thus, study of the $\Sigma^*$
resonance with available experimental data is necessary.

The $\Lambda p \to \Lambda p \pi^0$ reaction is a very good isospin
one filter for studying $\Sigma^*$ resonances decaying to $\pi
\Lambda$, and provides a useful tool for testing $\Sigma^*$ baryon
models. In the low energy region, the first $\Sigma(1193)$ excited
state, $\Sigma^*(1385)$, with strong couplings to $\pi \Lambda$
channel, should have significant contribution to the $\Lambda p \to
\Lambda p \pi^0$ reaction. The $\Sigma^*(1385)$ resonance is
cataloged in the baryon decuplet of the traditional quark models
that give a good description of the mass pattern and magnetic
moments for the baryon ground states. However, the classical quark
models still have some problems for the excited baryon resonances.
The lowest spatial excited states of baryon are expected to be a
$N^*$ ($uud$) state with one quark in orbital angular momentum $L =
1$ state, and hence should have negative parity. But,
experimentally, the lowest negative parity $N^*$ resonance is
$N^*(1535)$, which is heavier than $\Lambda(1405)$~\footnote{It is
worthy to mention that within the unitary chiral approaches, the
$N^*(1535)$ resonance and two $\Lambda(1405)$ states are dynamically
generated from the meson-baryon chiral
interaction~\cite{Inoue:2001ip,Jido:2003cb}.} and $N^*(1440)$ which
are spatial excited baryons. This is the long-standing mass reverse
problem for the lowest spatial excited baryons. Recently, the
penta-quark picture~\cite{Helminen:2000jb,Jaffe:2003sg} provides the
natural explanation for this problem~\cite{Zou:2007mk}. Based on the
penta-quark picture, a newly possible $\Sigma^*$ state,
$\Sigma^*(1380)$ ($J^P = 1/2^-$) was predicted around 1380
MeV~\cite{Zhang:2004xt}. Besides, another more general penta-quark
model~\cite{Helminen:2000jb} without introducing explicitly diquark
clusters also predicts this new $\Sigma^*$ state around 1405 MeV.
Obviously, it is helpful to check the correctness of penta-quark
models by studying the possible $\Sigma^*(1380)$ state. Because the
mass of this new $\Sigma^*$ state is close to the well established
$\Sigma^*(1385)$ resonance, it will make effects in the production
of $\Sigma^*(1385)$ resonance and then the analysis of the
$\Sigma^*(1385)$ resonance suffers from the overlapping mass
distributions and the common $\pi \Lambda$ decay mode. The possible
existence of such a new $\Sigma^*(1380)$ state in $J/\psi$ decays
was pointed out in Ref.~\cite{Zou:2006uh}. Recent studies on $K^- p
\to \Lambda \pi^+ \pi^-$ reaction have shown some evidence for the
existence of the $\Sigma^*(1380)$ state and width around $1380$ MeV
and $120$ MeV~\cite{Wu:2009tu,Wu:2009nw}. Furthermore, in
Refs.~\cite{Gao:2010hy,Chen:2013vxa}, the role played by the new
$\Sigma^*(1380)$ state in the $K \Sigma^*(1385)$ photo-production
reaction was studied, and it was shown that, apart from the existing
$\Sigma^*(1385)$ resonance, the $\Sigma^*(1380)$ state possibly
exists.

The $\Lambda p \to \Lambda p \pi^0$ reaction is difficult to study
experimentally because of the relatively small probability that the
short lived $\Lambda$ hyperon will interact with the target proton
rather than decay. Hence, little is known about this reaction. There
are only a few data points about its total cross section versus
energy~\cite{Kadyk:1971tc}, which was obtained in bubble chamber
measurements. The experimental results show a strong near threshold
enhancement. The $\Sigma^*(1385)$ resonance with spin-parity $3/2^+$
decays to $\pi \Lambda$ in relative $P$-wave and is suppressed at
low energies. To reproduce the near threshold enhancement for the
$\Lambda p \to \Lambda p \pi^0$ reaction, a natural source could be
some $J^P=1/2^-$, $\Sigma^*$ resonance(s) at low energy decay to
$\pi \Lambda$ in relative $S$-wave. Following the logic, in addition
to the $\Sigma^*(1385)$ resonance, we study the role played by the
possible $\Sigma^*(1380)$ state in the $\Lambda p \to \Lambda p
\pi^0$ reaction by using the effective Lagrangian method. The
production process is described by single-pion and single-kaon
exchange. Furthermore, the $\Lambda p$ final state interaction (FSI)
close to threshold is very strong and we also take it into account.
It is shown that the existence of the $\Sigma^*(1380)$ state can
also be tested in the $\Lambda p \to \Lambda p \pi^0$ reaction.

In the next section, we will show the formalism and ingredients in
our calculation, then numerical results and discussions are
presented in Sect. III. A short summary is given in the last
section.

\section{Formalism and ingredients}

The effective Lagrangian method is an important theoretical tool in
describing the various processes around the resonance region. But,
since only the tree diagrams are considered, thus the total
scattering amplitudes are not consistent with the unitary
requirements, which in principle is important for extracting the
parameters of the nucleon resonances from the analysis of the
experimental data~\cite{Kamano:2009im,Suzuki:2009nj}, especially for
those reactions involving many intermediate couple channels and
three-particle final states~\cite{Kamano:2008gr,Kamano:2011ih}. In
addition, it is known that it is difficult to really keep the
unitary in the three bodies case, which need to include the complex
loop diagrams~\cite{Kamano:2011ih,MartinezTorres:2009cw,alberto}.
Furthermore, the extracted rough parameters for the major resonances
still provide useful information, hence we will leave it to further
studies. Nevertheless, our model used in the present work can give a
reasonable description of the experimental data for the $\Lambda p
\to \Lambda p \pi^0$ reaction in the considered energy region, and
our calculation offers some important clues for the mechanisms of
the $\Lambda p \to \Lambda p \pi^0$ reaction and makes a first
effort to study the role of possible $\Sigma^*(1380)$ state in
relevant reaction.

In this section, we introduce the theoretical formalism and
ingredients to study the $\Lambda p \to \Lambda p \pi^0$ reaction by
using the effective Lagrangian method. In the following equations,
we use $\Sigma^*_1$ and $\Sigma^*_2$, which denote the
$\Sigma^*(1385)$ resonance and possible $\Sigma^*(1380)$ state,
respectively.

\subsection{Feynman diagrams and effective interaction Lagrangian densities} \label{feylag}

\begin{figure*}[htbp]
\begin{center}
\includegraphics[scale=0.8]{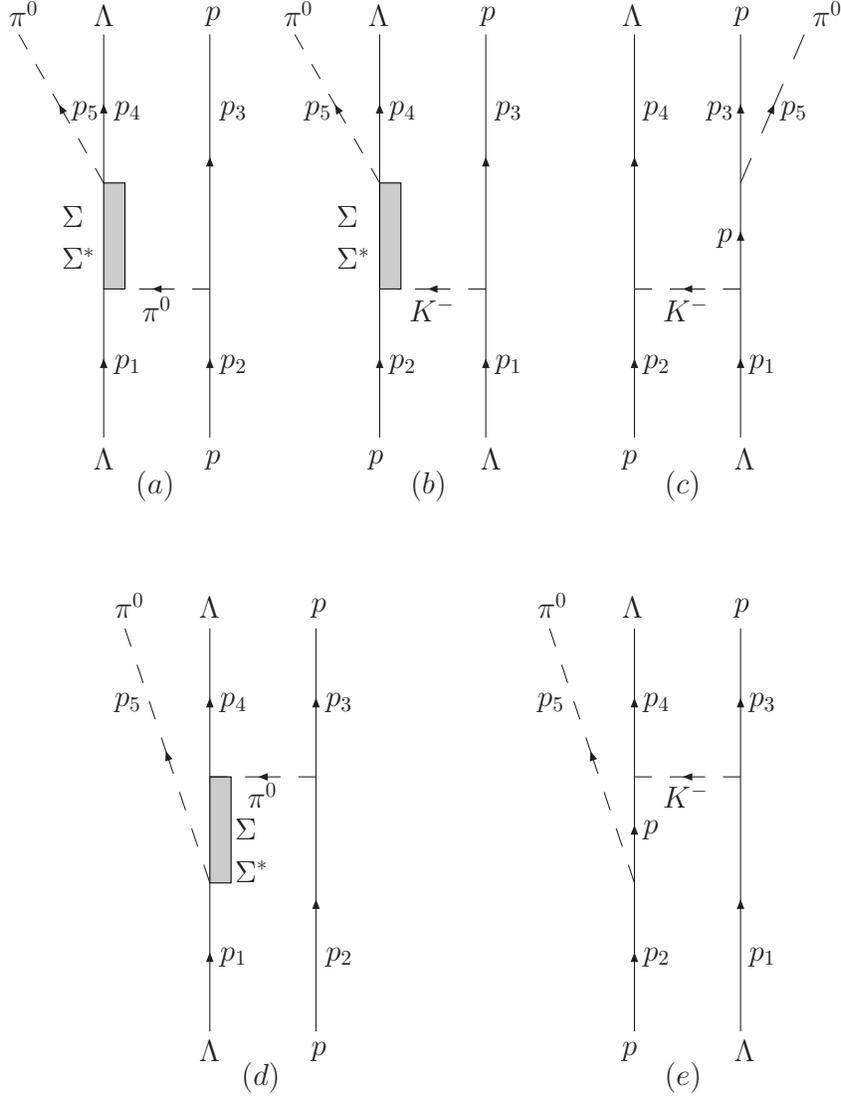}
\caption{Feynman diagrams for $\Lambda p \to \Lambda p \pi^0$
reaction.} \label{Fig:feyd}
\end{center}
\end{figure*}

To study the reaction of $\Lambda p \to \Lambda p \pi^0$, first we
investigate the possible reaction's mechanisms. In the reaction at
threshold, we consider the processes, shown in Fig.~\ref{Fig:feyd},
involving the exchange of $\pi$ [Fig.~\ref{Fig:feyd} (a), (d)] and
$K$ [Fig.~\ref{Fig:feyd} (b), (c) and (e)] mesons as the dominant
contributions. It is also assumed that the production of the $\pi^0
\Lambda$ passes mainly through the decay of the $\Sigma(1193)$,
$\Sigma^*(1385)$ and the possible $\Sigma^*(1380)$ state as shown in
Fig.~\ref{Fig:feyd} (a), (b) and (d). Besides, the contributions
from the nucleon pole are also considered as shown in
Fig.~\ref{Fig:feyd} (c) and (e).

To compute the contributions of those terms shown in
Fig.~\ref{Fig:feyd}, we use the interaction Lagrangian densities as
in
Refs.~\cite{Wu:2009tu,Wu:2009nw,Oh:2007jd,Gao:2012zh,Xie:2013wfa,Xie:2014kja},
\begin{eqnarray}
{\mathcal L}_{\pi N N}  &=& - \frac{g_{\pi NN}}{2m_N} \bar{N}
\gamma_5 \gamma_{\mu} \vec\tau
\cdot \partial^{\mu}\vec\pi N ,  \label{pinn} \\
{\mathcal L}_{K N \Lambda}  &=& - \frac{g_{K N
\Lambda}}{m_N+m_{\Lambda}} \bar{\Lambda}
\gamma_5 \gamma_{\mu} \partial^{\mu}K N \, +{\rm h.c.} \, ,  \label{klambdan} \\
{\mathcal L}_{\pi \Lambda \Sigma}  &=& - \frac{g_{\pi \Lambda
\Sigma}}{m_{\Lambda} + m_{\Sigma}} \bar{\Lambda}
\gamma_5 \gamma_{\mu} \partial^{\mu}\vec\pi \cdot \vec{\Sigma} \, +{\rm h.c.} \, ,  \label{pilambdasigma} \\
{\mathcal L}_{K N \Sigma}  &=& - \frac{g_{K N \Sigma}}{m_N +
m_{\Sigma}} \bar{N} \gamma_5 \gamma_{\mu} \partial^{\mu} K \vec\tau \cdot \vec{\Sigma} \, +{\rm h.c.} \, ,  \label{knsigma} \\
{\mathcal L}_{\pi \Lambda \Sigma^*_1} &=& \frac{g_{\pi \Lambda
\Sigma^*_1}}{m_{\pi}} \bar{\Sigma}^{* \mu}_1 (\vec\tau \cdot \partial_{\mu} \vec\pi) \Lambda \, +{\rm h.c.} \, ,  \label{pilambdasigmastar1} \\
\mathcal{L}_{K N \Sigma^*_1} &=& \frac{g_{K N \Sigma^*_1}}{m_{K}}
\bar{\Sigma}^{* \mu}_1 (\partial_{\mu} K) N \, +{\rm h.c.} \, ,
\label{knsigmastar1} \\
{\mathcal L}_{\pi \Lambda \Sigma^*_2} &=& g_{\pi \Lambda
\Sigma^*_2} \bar{\Sigma}^{*}_2 \vec\tau \cdot \vec\pi \Lambda \, +{\rm h.c.} \, ,  \label{pilambdasigmastar2} \\
\mathcal{L}_{K N \Sigma^*_2} &=& g_{K N \Sigma^*_2}
\bar{\Sigma}^{*}_2  K N \, +{\rm h.c.} \, , \label{knsigmastar2}
\end{eqnarray}
where $m_{\pi}$ and $m_K$ are the masses of pion and kaon,
respectively. The $\Sigma^{*\mu}_1$ and $\Sigma^*_2$ are the fields
for the $\Sigma^*(1385)$ resonance with spin-$\frac{3}{2}$ and
$\Sigma^*(1380)$ state with spin-$\frac{1}{2}$, respectively.

The coupling constant for $\pi NN$ vertex is taken to be $g_{\pi
NN}=13.45$, while the coupling constants $g_{KN\Lambda} = -13.98$,
$g_{\pi \Lambda \Sigma} = 9.32$ and $g_{KN\Sigma} = 2.69$ are
obtained from the ${\rm SU}(3)$ flavor symmetry. And these values
have also been used in previous
works~\cite{Xie:2013wfa,Xie:2014kja,Doring:2010ap,Xie:2013db} for
studying different processes.

For the coupling constant $g_{\pi \Lambda \Sigma^*(1385)}$, it can
be determined from the experimentally observed partial decay width
of $\Sigma^*(1385) \to \pi \Lambda$. With the effective interaction
Lagrangian described by Eq.~(\ref{pilambdasigmastar1}), the partial
decay width $\Gamma_{\Sigma^*(1385) \to \pi \Lambda}$ can be easily
calculated. The coupling constant are related to the partial decay
width as,~\footnote{With mass $M_{\Sigma^*(1385)} = 1384.57$ MeV,
total decay width $\Gamma_{\Sigma^*(1385)} = 37.13$ MeV and decay
branching ratio of $\Sigma^*(1385)$, Br[$\Sigma^*(1385) \to \pi
\Lambda$] = $0.87$, we obtain the coupling constant, $g_{\pi \Lambda
\Sigma^*_1} = 1.26$.}
\begin{eqnarray}
\Gamma_{\Sigma^*_1 \to \pi \Lambda} \! \! \! &=& \! \! \!
\frac{g^2_{\pi \Lambda \Sigma^*_1}}{12 \pi}
\frac{|\overrightarrow{p}^{{\rm c.m.}}_{\Lambda}|^3(E_{\Lambda} +
m_{\Lambda})}{m^2_{\pi} M_{\Sigma^*_1}}, \label{1385pilambda}
\end{eqnarray}
with
\begin{eqnarray}
E_{\Lambda}  &= & \frac{M^2_{\Sigma^*_1} + m^2_{\Lambda}
- m^2_{\pi}}{2M_{\Sigma^*_1}}, \\
|\overrightarrow{p}^{{\rm c.m.}}_{\Lambda}| &= & \sqrt{E^2_{\Lambda}
- m^2_{\Lambda}}.
\end{eqnarray}

For the $KN\Sigma^*(1385)$ coupling, it can be related with the $\pi
N \Delta$ coupling by the ${\rm SU(3)}$ flavor symmetry
relation~\cite{Oh:2007jd,Doring:2010ap}
\begin{eqnarray}
\frac{g_{\pi N \Delta}}{m_{\pi}} = -
\sqrt{6}\frac{g_{KN\Sigma^*_1}}{m_K}.
\end{eqnarray}

With the $\pi N \Delta$ coupling constant, $g_{\pi N \Delta} = 2.18$
obtained from the $\Delta$ decay width $\Gamma_{\Delta \to \pi N} =
120$ MeV~\footnote{With the Lagrangian, ${\mathcal L}_{\pi N \Delta}
= \frac{g_{\pi N \Delta}}{m_{\pi}} \bar{\Delta}^{\mu} (\vec\tau
\cdot \partial_{\mu} \vec\pi) N  +{\rm h.c.}$, we obtain for the
$\Delta \to \pi N$ decay width
\begin{eqnarray}
\Gamma_{\Delta \to \pi N} = \frac{g^2_{\pi N \Delta}}{12\pi
m^2_{\pi}} \frac{(E_N + m_N)(E_N^2 - m^2_N)^{3/2}}{M_{\Delta}},
\nonumber
\end{eqnarray}
where $E_N = \frac{M^2_{\Delta} + m^2_N -m^2_{\pi}}{2M_{\Delta}}$ is
the nucleon energy in the $\Delta$ rest frame.}, we obtain $g_{K
N\Sigma^*(1385)} = -3.19$ from the above equation.

Finally, we take the coupling constant $g_{\pi \Lambda
\Sigma^*(1380)}$ as $2.12$~\cite{Chen:2013vxa} which is obtained by
assuming the fitted results $120$ MeV of the $\Sigma^*(1380)$ total
decay width in Refs.~\cite{Wu:2009nw,Wu:2009tu} is contributed
totally by the $\pi \Lambda$ channel. On the other hand, for the
$KN\Sigma^*(1380)$ coupling, it is taken as $1.34$, which is the
fitted result of Refs.~\cite{Wu:2009nw,Wu:2009tu}.

In evaluating the scattering amplitudes of $\Lambda p \to \Lambda p
\pi^0$ reaction, we need to include the form factors because the
hadrons are not point like particles. We adopt here the common
scheme used in many previous works,
\begin{eqnarray}
F^{N N/\Lambda}_{\pi/K}(k^2_{\pi/K}) &=&
\frac{\Lambda^2_{\pi/K}-m_{\pi/K}^2}
{\Lambda^2_{\pi/K}- k_{\pi/K}^2} , \label{pinnff} \\
F^{\Lambda N /\Sigma}_{\pi/K}(k^2_{\pi/K}) &=&
\frac{\Lambda^2_{\pi/K}-m_{\pi/K}^2}
{\Lambda^2_{\pi/K}- k_{\pi/K}^2} , \label{pisigmalambda} \\
F^{\Sigma^* \Lambda/N}_{\pi/K}(k^2_{\pi/K}) &=&
\left(\frac{\Lambda^{*2}_{\pi/K}-m_{\pi/K}^2}{\Lambda^{*2}_{\pi/K}-k_{\pi/K}^2}\right)^n,
\label{pinnstarff} \\
F_{\Sigma^*}(q^2_{\Sigma^*}) &=& \left[
\frac{\Lambda^4_{\Sigma^*}}{\Lambda^4_{\Sigma^*} + (q^2_{\Sigma^*} -
M^2_{\Sigma^*})^2} \right]^n , \\
F_{p}(q^2_{p}) &=& \frac{\Lambda^4_{p}}{\Lambda^4_{p} + (q^2_{p} -
m^2_{p})^2} , \\
F_{\Sigma}(q^2_{\Sigma}) &=&
\frac{\Lambda^4_{\Sigma}}{\Lambda^4_{\Sigma} + (q^2_{\Sigma} -
m^2_{\Sigma})^2} ,
\end{eqnarray}
where $k_{\pi} = p_2 - p_3$ [Fig.~\ref{Fig:feyd} (a), (d)], $k_K =
p_1 - p_3$ [Fig.~\ref{Fig:feyd} (b), (e)], $k_K = p_4 - p_2$
[Fig.~\ref{Fig:feyd} (c)] are the 4-momentum of the exchanged
$\pi^0$ meson, $K^-$ meson, while $q_{\Sigma/\Sigma^*} = p_4 + p_5$
[Fig.~\ref{Fig:feyd} (a), (b)], $q_{\Sigma/\Sigma^*} = p_1 - p_5$
[Fig.~\ref{Fig:feyd} (d)] and $q_p = p_3 + p_5$ [Fig.~\ref{Fig:feyd}
(c)], $q_p = p_2 -p_5$ [Fig.~\ref{Fig:feyd} (e)] are the 4-momentum
of the $\Sigma^*$ resonances and the nucleon pole. On the other
hand, we take $n=2$ for $\Sigma^*(1385)$ resonance and $n=1$ for
$\Sigma^*(1380)$ state. The $\Lambda_{\pi/K}$, $\Lambda^*_{\pi/K}$
and $\Lambda_{\Sigma^*}$ are cut-off parameters, which are taken as
commonly used ones: $\Lambda_{\pi} = \Lambda_{K} = 1.3$ GeV,
$\Lambda^{*}_{\pi} = \Lambda^{*}_{K} = \Lambda_{\Sigma^*} =
\Lambda_p = \Lambda_{\Sigma} = 0.8$ GeV.

\subsection{Scattering amplitudes}

To get the invariant scattering amplitudes for the reaction $\Lambda
p \to \Lambda p \pi^0$, we need also the propagators for $\pi$ and
$K$ mesons, nucleon pole, $\Sigma(1193)$ pole, $\Sigma^*(1380)$
state and $\Sigma^*(1385)$ resonances~\footnote{It is worthy to note
that we take $\Gamma_{\Sigma^*_{1,2}} = 0$ for calculation of
Fig.~\ref{Fig:feyd} (d) since $q^2_{\Sigma^*_{1,2}} < 0$ in this
case.},
\begin{eqnarray}
G_{\pi/K}(k^2_{\pi/K}) &=&  \frac{i}{k^2_{\pi/K}-m^2_{\pi/K}}, \\
G_{p}(q_p) &=& i \frac{\Slash q_{p} +m_{p}}{q^2_{p} - m^2_{p} }, \\
G_{\Sigma}(q_{\Sigma}) &=& i \frac{\Slash q_{\Sigma}
+m_{\Sigma}}{q^2_{\Sigma} - m^2_{\Sigma} },
\end{eqnarray}

\begin{eqnarray}
G_{\Sigma^*_2}(q_{\Sigma^*_2}) &=& i \frac{\Slash q_{\Sigma^*_2} + M_{\Sigma^*_2}}{q^2_{\Sigma^*_2} - M^2_{\Sigma^*_2} + i M_{\Sigma^*_2} \Gamma_{\Sigma^*_2}}, \\
G^{\mu\nu}_{\Sigma^*_1}(q_{\Sigma^*_1}) &=& i \frac{\Slash
q_{\Sigma^*_1} + M_{\Sigma^*_1}}{D}P^{\mu\nu},
\end{eqnarray}
with
\begin{eqnarray}
D &=& s-M^2_{\Sigma^*_1} + i M_{\Sigma^*_1}\Gamma_{\Sigma^*_1}, \\
P^{\mu \nu} &=& -g^{\mu \nu} + \frac{1}{3}\gamma^{\mu}\gamma^{\nu} +
\frac{2}{3M^2_{\Sigma^*_1}}q_{\Sigma^*_1}^{\mu}q_{\Sigma^*_1}^{\nu} \nonumber \\
&& +
\frac{1}{3M_{\Sigma^*_1}}(\gamma^{\mu}q_{\Sigma^*_1}^{\nu}-\gamma^{\nu}q_{\Sigma^*_1}^{\mu}),
\end{eqnarray}
where $M_{\Sigma^*_1}$ ($M_{\Sigma^*_2}$) and $\Gamma_{\Sigma^*_1}$
($\Gamma_{\Sigma^*_2}$) are the mass and total decay width of the
$\Sigma^*(1385)$ [$\Sigma^*(1380)$] resonance, respectively. We take
$M_{\Sigma^*_2}$ and $\Gamma_{\Sigma^*_2}$ as $1380$ MeV and $120$
MeV which were used in Refs.~\cite{Wu:2009tu,Wu:2009nw}.

Then, the full invariant scattering amplitude of the $\Lambda p \to
\Lambda p \pi^0$ reaction is composed of five parts corresponding to
the diagrams shown in Fig.~\ref{Fig:feyd},
\begin{eqnarray}
{\cal M} &=& {\cal M}^{\Sigma, \Sigma^*_1, \Sigma^*_2}_{a} + {\cal
M}^{\Sigma, \Sigma^*_1, \Sigma^*_2}_{b} + {\cal M}^p_{c} + \nonumber
\\
&& {\cal M}^{\Sigma, \Sigma^*_1, \Sigma^*_2}_{d} + {\cal M}^p_{e}.
\label{ppamp}
\end{eqnarray}

Each of the above amplitudes can be obtained straightforwardly with
the effective couplings and following the Feynman rules. Here we
give explicitly the amplitude ${\cal M}_{a}$ for the
$\Sigma^*(1380)$ state, as an example,
\begin{eqnarray}
{\cal M}^{\Sigma^*_2}_{a} & = & g_{\pi NN} g^2_{\pi \Lambda
\Sigma^*_2} F^{N N}_{\pi}(k^2_{\pi}) F^{\Sigma^*_2
\Lambda}_{\pi}(k^2_{\pi}) F_{\Sigma^*_2}(q_{\Delta^*}^2) \nonumber\\
&&  \times  \bar{u} (p_4,s_4) G_{\Sigma^*_2}(q_{\Sigma^*_2}) u(p_1,s_1) G_{\pi}(k^2_{\pi})  \nonumber\\
&&  \times \bar{u}(p_3,s_3) \gamma_5 u(p_2,s_2) , \label{ppampms}
\end{eqnarray}
where $s_i~(i=1,2,3,4)$ and $p_i~(i=1,2,3,4)$ represent the spin
projection and 4-momenta of the initial and final $\Lambda$ hyperons
and protons, respectively.

\subsection{Final state interaction}

To study possible influence from the $\Lambda p$ FSI, we include it
in our calculation by introducing a FSI enhancement factor $|C_{\rm
FSI}|^2$,
\begin{eqnarray}
|{\cal M}|^2 \rightarrow |{\cal M}|^2|C_{\rm FSI}|^2,
\end{eqnarray}
where the correction $C_{\rm FSI}$ is given as
\begin{eqnarray}
C_{\rm FSI} = \frac{q - i\beta}{q + i \alpha}
\end{eqnarray}
where $q$ is the internal momentum of $\Lambda p$ subsystem, and the
$\alpha$ and $\beta$ are related to the spin-averaged scattering
lengths $\bar{a}$ and the effective ranges $\bar{r}$ of the low
energy $S$-wave scattering,
\begin{eqnarray}
\alpha =
\frac{1}{\bar{r}}\left(1-\sqrt{1-\frac{2\bar{r}}{\bar{a}}}\right),
~~\beta =
\frac{1}{\bar{r}}\left(1+\sqrt{1-\frac{2\bar{r}}{\bar{a}}}\right).
\end{eqnarray}
with $\bar{a} = -1.75$ and $\bar{r} = 3.43$ obtained in
Refs.~\cite{Hinterberger:2004ra,Sibirtsev:2005mv,Sibirtsev:2006uy,Xie:2011me},
we get $\alpha = -70.1$ MeV and $\beta = 185.1$ MeV.

To end this section, it is worth to mention that, in general, the
$\Lambda p$ interaction is spin-dependent. Thus, to analyze the low
energy elastic $\Lambda p \to \Lambda p$ transition cross section,
we need four parameters: scattering length $a_s$ and effective range
$r_s$ for the spin of $\Lambda p$ system $S_{\Lambda p} = 0$;
scattering length $a_t$ and effective range $r_t$ for $S_{\Lambda p}
= 1$. However, the current lower energy experimental data on the
$\Lambda p \to \Lambda p$ reaction and $pp \to \Lambda p K^+$
reaction only support the determination of a spin-averaged
scattering length $\bar{a}$ and effective range
$\bar{r}$~\cite{Hinterberger:2004ra,Sibirtsev:2005mv,Sibirtsev:2006uy,Xie:2011me}.
Indeed, as pointed in Ref.~\cite{Xie:2011me}, only two parameters in
the $\Lambda p$ interaction are enough to reproduce the current
experimental data on low energy $\Lambda p$ scattering.

\section{Numerical results and discussion}

With the formalism and ingredients given above, the calculations of
the differential and total cross sections for $\Lambda p \to \Lambda
p \pi^0$ are straightforward,
\begin{eqnarray}
&& d\sigma (\Lambda p \to \Lambda p \pi^0) =
\frac{1}{4}\frac{m_{\Lambda} m_p}{F}
\sum_{s_1, s_2} \sum_{s_3, s_4} |{\cal M}|^2 \times \nonumber \\
&& \frac{m_p d^{3} p_{3}}{E_{3}}
\frac{m_{\Lambda} d^{3} p_4}{E_4} \frac{d^{3} p_5}{2 E_5} \delta^4 (p_{1}+p_{2}-p_{3}-p_{4}-p_5), \nonumber \\
\label{ppdcs}
\end{eqnarray}
with the flux factor
\begin{eqnarray}
F=(2 \pi)^5\sqrt{(p_1\cdot p_2)^2-m^2_{\Lambda} m^2_p}~.
\label{eqff}
\end{eqnarray}

\begin{figure}[htbp]
\begin{center}
\includegraphics[scale=0.45]{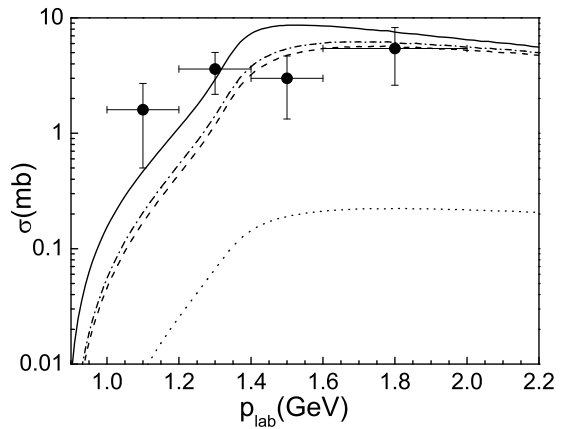}
\caption{Total cross sections vs the beam momentum p$_{\rm lab}$ for
$\Lambda p \to \Lambda p \pi^0$ reaction. The experimental data are
taken from Ref.~\cite{Kadyk:1971tc}. The dashed and dotted curves
stand the contributions of $\pi^0$ and $K^{-}$ exchange,
respectively, while the sold (with $\Lambda p$ FSI) and dash-dotted
(without $\Lambda p$ FSI) are their total contributions.}
\label{Fig:tcspionkaon}
\end{center}
\end{figure}

The theoretical results of the total cross section for beam energies
${\rm p_{lab}}$ from just above the production threshold $0.9$ GeV
to $2.2$ GeV are shown in Fig.~\ref{Fig:tcspionkaon}. In this figure
we have investigated the role of various meson exchange processes in
describing the total cross section. The dashed and dotted lines
stand for contributions from $\pi^0$ and $K^-$ exchange,
respectively. Their total contributions are shown by the dash-dotted
line, while the results with the $\Lambda p$ FSI are shown by the
solid line. It is found that $\Lambda p$ FSI enhance the total cross
section by a factor $3$ close to reaction threshold. Thus the
$\Lambda p$ FSI is indeed making a significant effect at very low
energies. But it does not change the basic shape of the curve very
much. Besides, from Fig.~\ref{Fig:tcspionkaon}, we can see that the
contribution from $\pi^0$ meson exchange is predominant in the whole
considered energy region, and the contribution from the $K^-$ meson
exchange is rather small and can be negligible. For comparison, we
also show the experimental data~\cite{Kadyk:1971tc} in
Fig.~~\ref{Fig:tcspionkaon}, from where we can see that the measured
total cross sections are reproduced reasonably well by our model
calculations (solid line).

\begin{figure}[htbp]
\begin{center}
\includegraphics[scale=0.45]{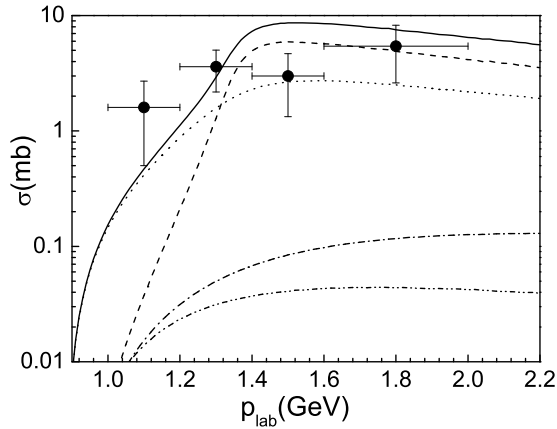}
\caption{Contributions of $\Sigma^*(1385)$ resonance (dashed line),
$\Sigma^*(1380)$ state (dotted line), nucleon pole (dash-dotted
line) and $\Sigma(1193)$ pole (dash-dot-dotted line) to the total
cross sections vs the beam momentum p$_{\rm lab}$ for the $\Lambda p
\to \Lambda p \pi^0$ reaction. Their total contribution is shown by
solid line. The experimental data are taken from
Ref.~\cite{Kadyk:1971tc}.} \label{Fig:tcsres}
\end{center}
\end{figure}

The relative importance of the contributions of each intermediate
resonance to the $\Lambda p \to \Lambda p \pi^0$ reaction is studied
in Fig.~\ref{Fig:tcsres}, where the contributions of
$\Sigma^*(1385)$ resonance, $\Sigma^*(1380)$ state, nucleon pole and
$\Sigma(1193)$ pole to the energy dependence of the total cross
section are shown by dashed, dotted, dash-dotted, and
dash-dot-dotted curves, respectively. Their total contribution is
depicted by the solid line. It is clear that the contributions from
the $\Sigma^*(1380)$ state and $\Sigma^*(1385)$ resonance dominate
the total cross section at beam momenta below and above $1.3$ GeV,
respectively, while the contributions of nucleon and $\Sigma(1193)$
pole are small and can be neglected.

As mentioned in the introduction, the $\Sigma^*(1385)$ resonance
with spin-parity $3/2^+$ decays to $\pi \Lambda$ in relative
$P$-wave and is suppressed at low energies. It can not reproduce the
near threshold enhancement for the $\Lambda p \to \Lambda p \pi^0$
reaction. On the contrary, the possible $\Sigma^*(1380)$ state with
$J^P=1/2^-$ is decaying to $\pi \Lambda$ in relative $S$-wave, which
will give enhancement at the near threshold. As we can see in
Fig.~\ref{Fig:tcsres}, thanks to the contribution from
$\Sigma^*(1380)$ state, we can reproduce the experimental data for
all of the beam energies. Thus, we find a natural source for the
near threshold enhancement of the $\Lambda p \to \Lambda p \pi^0$
reaction coming from the possible $\Sigma^*(1380)$ state which
decays to $\pi \Lambda$ in the $S$-wave.

In addition to the total cross sections, we also compute the
differential cross sections for $\Lambda p \to \Lambda p \pi^0$
reaction, namely the angular distributions of all final-state
particles in the overall center-of-mass frame (CMS), as well as
distributions in both the Gottfried-Jackson and helicity frames as
introduced in Refs.~\cite{AbdelBary:2010pc}. Like Dalitz plots, the
helicity angle distributions provide insight into the three-body
final state. While the information contained in the
Gottfried-Jackson angle distributions is complementary to that of a
Dalitz plot, as this angular distribution can give insight into the
scattering process, especially concerning the involved partial
waves.

\begin{figure*}[htbp]
\includegraphics[scale=0.8]{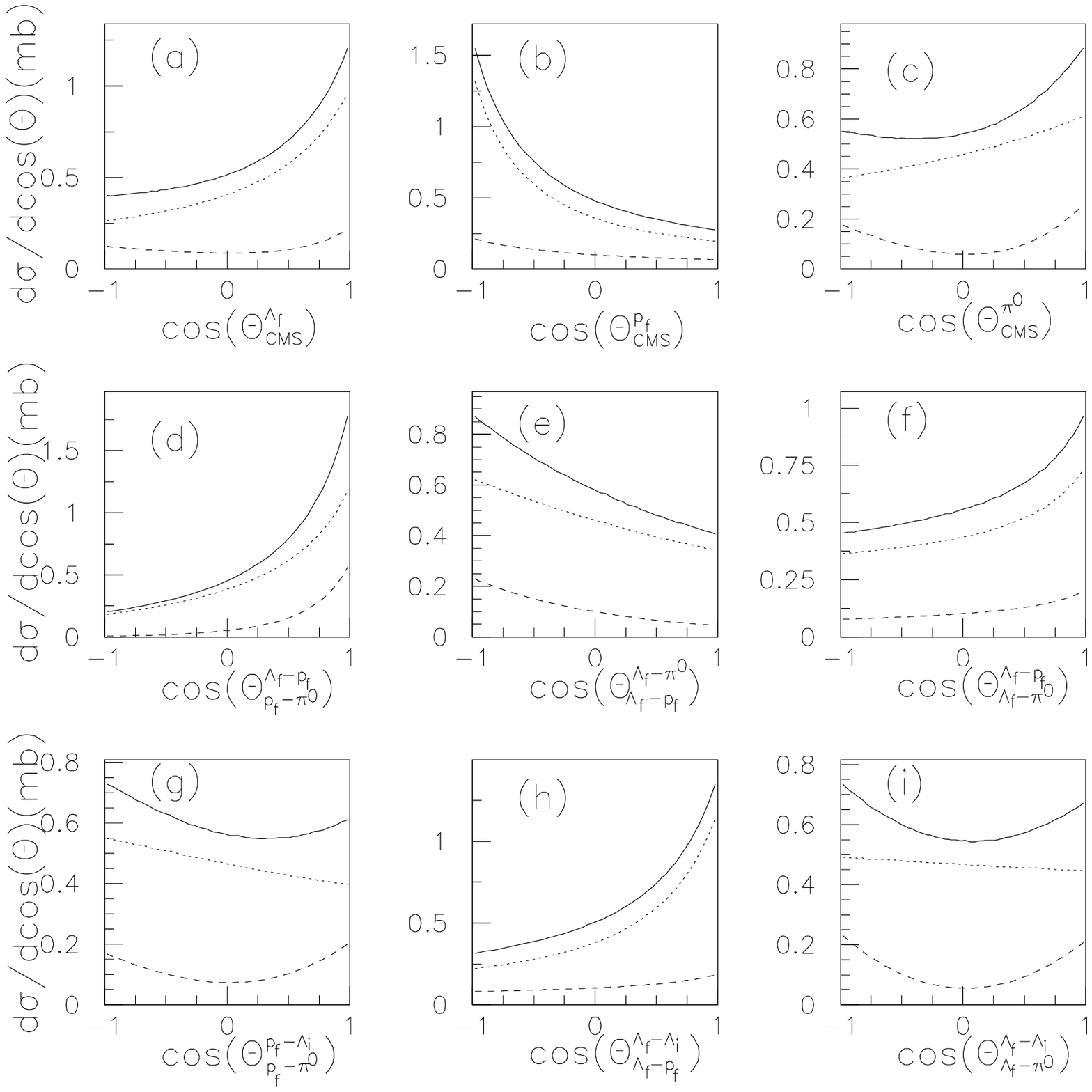}
\vspace{-0.2cm} \caption{Angular differential cross sections for the
$\Lambda p \to \Lambda p \pi^0$ reaction in CMS [(a):
$\Theta^{\Lambda_{\rm f}}_{\rm CMS}$, (b): $\Theta^{p_{\rm f}}_{\rm
CMS}$, (c): $\Theta^{\pi^0}_{\rm CMS}$], helicity [(d):
$\Theta^{\Lambda_{\rm f}-p_{\rm f}}_{p_{\rm f}-\pi^0}$, (e):
$\Theta^{\Lambda_{\rm f}-\pi^0}_{\Lambda_{\rm f}-p_{\rm f}}$, (f):
$\Theta^{\Lambda_{\rm f}-p_{\rm f}}_{\Lambda_{\rm f}-\pi^0}$], and
Gottfried-Jackson [(g): $\Theta^{p_{\rm f}-\Lambda_{\rm i}}_{p_{\rm
f}-\pi^0}$, (h): $\Theta^{\Lambda_{\rm f}-\Lambda_{\rm
i}}_{\Lambda_{\rm f}-p_{\rm f}}$, (i): $\Theta^{\Lambda_{\rm
f}-\Lambda_{\rm i}}_{\Lambda_{\rm f}-\pi^0}$] reference frames. The
dashed and solid curves stand the contributions of the
$\Sigma^*(1385)$ and $\Sigma^*(1380)$, respectively. The results are
obtained at p$_{\rm lab} = 1.2$ GeV.} \label{Fig:angdis12}
\end{figure*}

\begin{figure*}[htbp]
\includegraphics[scale=0.8]{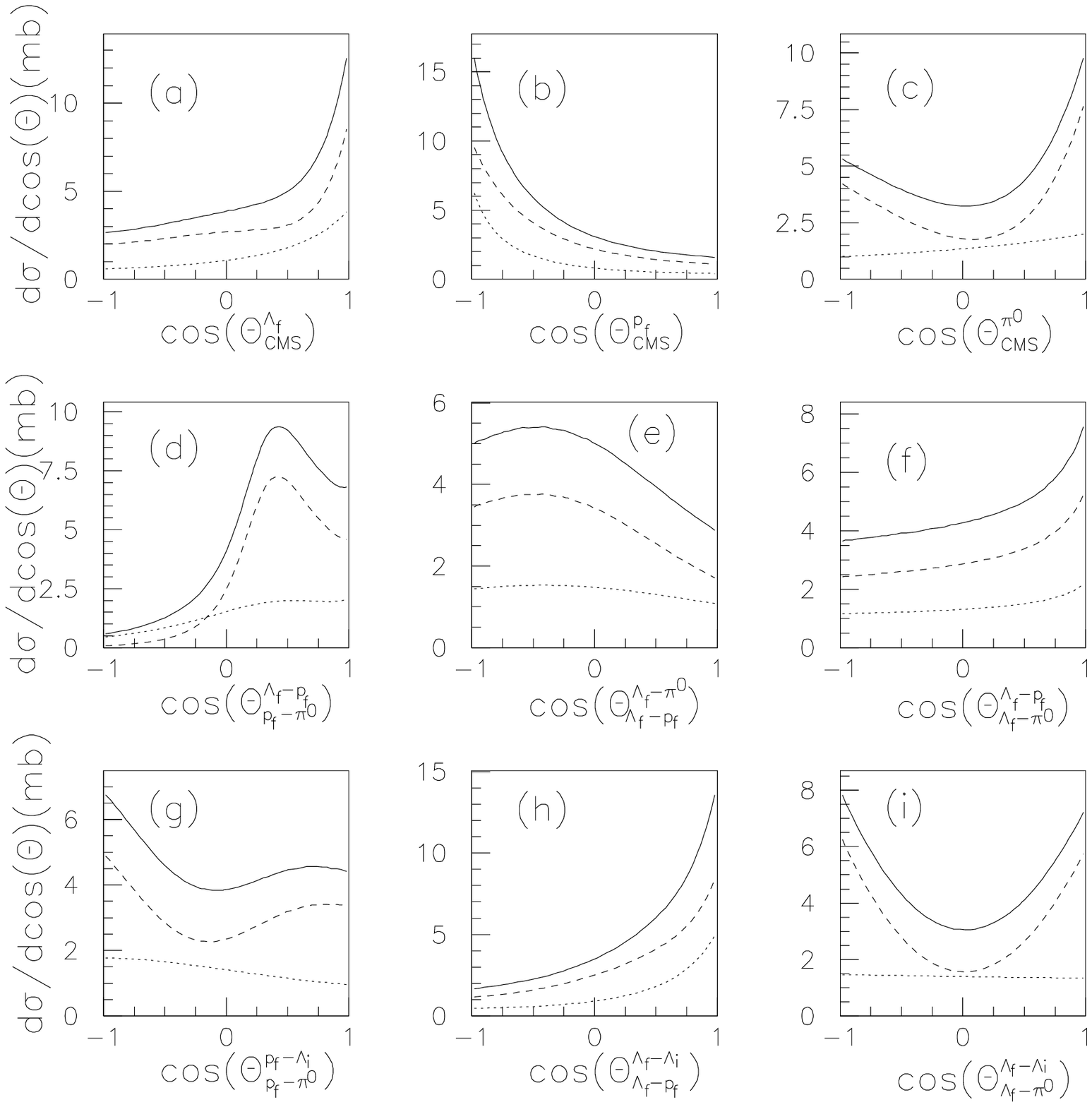}
\vspace{-0.2cm} \caption{As in Fig.~\ref{Fig:angdis12} but for the
case of p$_{\rm lab} = 1.5$ GeV.} \label{Fig:angdis15}
\end{figure*}

The corresponding theoretical results at p$_{\rm lab} = 1.2$ GeV,
where the contribution of $\Sigma^*(1380)$ state is dominant, are
shown in Fig.~\ref{Fig:angdis12}.~\footnote{The $\Lambda_{\rm i}$
($\Lambda_{\rm f}$) and p$_{\rm i}$ (p$_{\rm f}$) stand for the
initial (final) $\Lambda$ hyperon and proton, respectively.} For
comparison, we also show our theoretical predictions in
Fig.~\ref{Fig:angdis15} at p$_{\rm lab} = 1.5$ GeV, where the
contribution of $\Sigma^*(1385)$ resonance is dominant. In those
figures, the dashed and dotted curves are obtained with the
contributions from $\Sigma^*(1385)$ resonance and $\Sigma^*(1380)$
state, respectively. The solid lines stand for their total
contributions.

In Figs.~\ref{Fig:angdis12}, \ref{Fig:angdis15} (a), (b), and (c),
we show the final particles $\Lambda$, $p$ and $\pi^0$ angular
distributions in the CMS, respectively. The results obtained in the
helicity frame with respect to the angle, $\Theta^{a-b}_{c-d}$,
which represents the angel between particles ``$a$" and ``$b$" in
the ``$c$" and ``$d$" reference frame (see more details in
Refs.~\cite{AbdelBary:2010pc,Agakishiev:2011qw}), are shown in
Figs.~\ref{Fig:angdis12}, \ref{Fig:angdis15} (d), (e), and (f),
while Figs.~\ref{Fig:angdis12}, \ref{Fig:angdis15} (g), (h), and (i)
depict the distributions of the Gottfried-Jackson angles. It is
worthy to mention that the nine angular distributions are not
kinematically independent with each other, we show here all of them
for the sake of completeness.

From Figs.~\ref{Fig:angdis12}, \ref{Fig:angdis15}, we can see that
the shapes of the angular distributions of $\Sigma^*(1385)$
resonance and $\Sigma^*(1380)$ state are much different, so the
existence of $\Sigma^*(1380)$ state can be tested by future
experimental analysis.

\section{Summary}

In summary, we study the $\Lambda p \to \Lambda p \pi^0$ reaction
near threshold within an effective Lagrangian method. In addition to
the role played by the $\Sigma^*(1385)$ resonance (spin-parity $J^P
= 3/2^+$), we study the effects of a newly proposed $\Sigma^*$ ($J^P
= 1/2^-$) state with mass and width around $1380$ MeV and $120$ MeV.
We show that our model leads to a fair description of the
experimental data on the total cross section of the $\Lambda p \to
\Lambda p \pi^0$ reaction by including the contributions from the
possible $\Sigma^*(\frac{1}{2}^-)$ state and the strong $\Lambda p$
FSI.

The $\Sigma^*(1385)$ resonance can not reproduce the near threshold
enhancement for the $\Lambda p \to \Lambda p \pi^0$ reaction because
it decays to $\pi \Lambda$ in relative $P$-wave and is suppressed at
low energies. On the contrary, the newly $\Sigma^*(1380)$ state
decays to $\pi \Lambda$ in relative $S$-wave, and can describe the
near threshold enhancement fairly well, which indicate that the
$\Lambda p \to \Lambda p \pi^0$ data support the existence of this
$\Sigma^*(1380)$ state, and more accurate data for this reaction can
be used to improve our knowledge on the $\Sigma^*(1380)$ properties.
Our present calculation offers some important clues for the
mechanisms of the $\Lambda p \to \Lambda p \pi^0$ reactio and makes
a first effort to study the role of the $\Sigma^*(1380)$ state in
relevant reaction.

\section*{Acknowledgments}

We would like to thank Prof. T.-S.H. Lee and Xu Cao for useful
discussions. This work is partly supported by the National Natural
Science Foundation of China under grants 11105126, 11035006,
11121092, 11261130311 (CRC110 by DFG and NSFC), the Chinese Academy
of Sciences under Project No. KJCX2-EW-N01 and the Ministry of
Science and Technology of China (2009CB825200). This material is
based upon work supported by the U.S. Department of Energy, Office
of Science, Office of Nuclear Physics, under contract number
DE-AC02-06CH11357.

\end{document}